# Photonic microwave signals with zeptosecond level absolute timing noise

Xiaopeng Xie[1]★, Romain Bouchand[1]★, Daniele Nicolodi[1]#, Michele Giunta[2,6], Wolfgang Hänsel[2], Matthias Lezius[2], Abhay Joshi[3], Shubhashish Datta[3], Christophe Alexandre[4], Michel Lours[1], Pierre-Alain Tremblin[5], Giorgio Santarelli[5], Ronald Holzwarth[2,6] and Yann Le Coq[1]★

[1]LNE-SYRTE, Observatoire de Paris, PSL Research University, CNRS, Sorbonne Universités, UPMC Univ. Paris 06, 61 avenue de l'Observatoire, 75014 Paris, France
[2] Menlo Systems GmbH, Am Klopferspitz 19a, D-82152 Martinsried, Germany
[3]Discovery Semiconductors Inc., 119 Silvia Street, Ewing, New Jersey 08628, USA
[4]CNAM, CEDRIC Laboratory, 292 rue Saint Martin, 75003 Paris, France
[5]LP2N, IOGS - CNRS - Université de Bordeaux 1, rue F. Mitterand, 33400 Talence, France
[6]Max-Planck-Institut für Quantenoptik, 85748 Garching, Germany
★These authors contributed equally to this work
#Present address: National Institute of Standards and Technology, Boulder, Colorado 80305, USA
★E-mail: yann.lecoq@obspm.fr

**Photonic synthesis of radiofrequency revived the quest for unrivalled microwave purity by its seducing ability to convey the benefits of the optics to the microwave world[1-11]. In this work, we perform a high-fidelity transfer of frequency stability between an optical reference and a microwave signal via a low-noise fiber-based frequency comb and cutting-edge photo-detection techniques. We demonstrate the generation of the purest microwave signal with a fractional frequency stability below 6.5 x 10$^{-16}$ at 1 s and a timing noise floor below 41 zs.Hz$^{-1/2}$ (phase noise below -173 dBc.Hz$^{-1}$ for a 12 GHz carrier). This outclasses existing sources and promises a new era for state-of-the-art microwave generation. The characterization is achieved through a heterodyne cross-correlation**

**scheme with lowermost detection noise. This unprecedented level of purity can impact domains such as radar systems[12], telecommunications[13] and time-frequency metrology[2,14]. The measurement methods developed here can benefit the characterization of a broad range of signals.**

Photonic microwave generation methods, like optoelectronic oscillator[4], Brillouin oscillator[5], sideband-injection-locked laser[6], electro-optical frequency division[7], and Kerr frequency comb oscillator[8] have drawn the attention for their interesting properties such as high frequency, large bandwidth, tunability, chip-scale packaging. In particular, ultra-stable lasers and optical frequency comb based optical-to-radio frequency division scheme can produce microwave signals with both extremely high stability and low noise[2,3,9-11]. In this work, we advance this approach to an unprecedented level.

Optical frequency combs are lasers emitting a series of evenly spaced ultra-short pulses, corresponding to discrete equidistant lines in the optical spectrum[15,16]. The frequency of the comb lines $\nu_N$ can be expressed as $\nu_N = Nf_r + f_0$, where N is an integer number, $f_r$ is the pulse repetition rate and $f_0$ is the carrier envelope offset (CEO) frequency. In our setup, one of the comb lines is tightly phase locked to a narrow linewidth continuous-wave (CW) laser reference $\nu_{CW}$ exhibiting very high stability and spectral purity. In addition, the CEO frequency $f_0$ is measured with the f-2f self-referencing technique, and removed in the phase lock process to obtain a virtual "$f_0$-free" comb obeying $v_N - f_0 = Nf_r = v_{CW}$ [2,3,17]. The spectrum of the electrical output of a photodiode illuminated by the comb light consists of harmonics of the repetition rate up to the cutoff frequency of the detector. Any aforementioned harmonics of order n can then be isolated to yield a microwave signal $f_\mu = nf_r$ whose absolute frequency stability is equal to that of the reference light. Accordingly, the stability is said to be transferred from the optics to the microwaves: $\delta v_{CW}/v_{CW} = \delta f_r/f_r = \delta f_\mu/f_\mu$. Thanks to the carrier frequency division, the phase noise power spectral density is intrinsically reduced by $M^2$, where $M = v_{CW}/f_\mu = N/n \sim 10^4$ is the frequency division factor.

The microwave generation system is sketched in Fig.1: a low-noise Er-doped fiber-based optical frequency comb featuring 250 MHz repetition rate acts as a frequency divider; an ultra-stable CW laser at 1542 nm, with a fractional frequency stability as low as 5.5 x $10^{-16}$ at 1 s[18], is used as a reference for stabilizing the comb; a specially designed high-linearity photodiode with flicker phase noise below -140 $f^{-1}$ dBc.Hz$^{-1}$ is used to convert the optical trains of pulses to a microwave signal.

During the optical-to-microwave conversion process, sundry important techniques are proposed and implemented. To increase the signal-to-noise ratio, a fiber-based optical pulse repetition rate multiplier is used to redistribute the photocurrent to the harmonic of interest[19] (the 48th in our case). After four multiplying stages, we obtain over 2.5 mW of microwave power at 12 GHz, for a DC photocurrent of 8 mA. A dispersion compensation fiber unit is inserted for ensuring the optical pulses impinging on the photodiode are less than 800 fs-long (cf. Methods). In this ultrashort pulse limit, the shot-noise distribution imbalance between amplitude and phase allows to reduce the shot-noise-induced phase noise[20]. Careful pulse compression proved essential for achieving the final result.

Fiber-based lasers exhibit significant relative intensity noise (RIN) which can result in excess phase noise through amplitude-to-phase conversion in the photodiode. We limit the RIN by meticulously choosing the laser working state (*cf.* Methods) and operate the photodiode in conditions where non-linear saturation effects help zeroing the amplitude-to-phase (AM-PM) conversion for the 12 GHz signal[21]. Rejection of amplitude noise by more than 33 dB is typically obtained.

Combining these forefront opto-electronics devices and cutting-edge techniques, microwave signals with phase noise below -173 dBc.Hz$^{-1}$ at 10 kHz offset are made possible. As phase noise is only a reasonable figure of merit of an oscillator provided its carrier frequency, we choose to express it as a timing noise density, once normalized against the 12 GHz carrier. It represents an impressive timing noise below 41 zs.Hz$^{-1/2}$.

The thorough characterization of such ultra-low phase noise signals is a major challenge in itself, even more so for high-frequency carriers. In high-precision measurements, phase noise characterization is usually a process that involves comparing a signal from the device under test (DUT) with a reference source. When the signal under test has lower phase noise than the available reference, two separate but identical systems can be built and compared[9]. The data are then analyzed assuming that the two identical systems contribute equally to the phase noise. However, realizing two equally good systems is not straightforward and minute excess phase noise of the phase comparison system can strongly impact the final result obtained by such methods.

To overcome these limitations, we demonstrate a heterodyne digital cross-correlation scheme based on three similar but independent optoelectronic microwave generation systems, as sketched in Fig.1. By frequency mixing the DUT 12 GHz microwave signal with signals from the two auxiliary systems, two beat-notes around 5 MHz are obtained. These beat-note signals, both carrying phase information from the DUT, are sent to an FPGA-based heterodyne cross-correlator where they are sampled by fast analog-to-digital converters, digitally down-converted, and processed to generate two independent phase-comparison data sets. Cross-correlation of these two phase noise data sets converges to the absolute phase noise power spectral density (PSD) of the 12 GHz signal that we want to characterize. The microwave signals generated by the two auxiliary systems act as phase references but do not need to be as good as the signal being characterized. Their uncorrelated noise determines the uncertainty of the estimates of the phase noise PSD. This uncertainty scales as $1/\sqrt{m}$, with $m$ the number of averages, and can thus be made arbitrarily small in the assumption of stationary DUT phase noise[22]. The measurement noise floor is below -180 dBc.Hz$^{-1}$ for Fourier frequencies beyond 1 kHz offset, see Fig. 2 (black curve).

To assess the additional phase noise introduced by the optical frequency division process, including comb and photodetectors, we measure the microwave phase noise obtained when locking the two auxiliary microwave references and the DUT to the same ultra-stable optical reference. In this configuration, the laser noise is common to the three systems and then suppressed. The obtained phase noise PSD is shown in Fig.2 (red curve). At Fourier frequencies up to 3 kHz, although there are numerous peaks due to acoustic noise, the photodiode flicker noise can be inferred to be below -140 f$^{-1}$ dBc.Hz$^{-1}$, which is the lowest level ever reported to our knowledge[23]. At lower Fourier frequencies, the phase noise departs from a 1/f behavior, which we explain by residual long-term drift due to the noisy laboratory environment and the significant amount of non-noise-cancelled fibered optics (several tenths of meters mostly in the pulse rate multipliers). From 3 kHz to 100 kHz, we measure a shot-noise-limited white phase noise floor at -173 dBc.Hz$^{-1}$ that is more than 10 dB below previously reported results obtained with comparable schemes[9,24]. This shot noise floor corresponds to a timing noise of 41 zs.Hz$^{-1/2}$ and depends on the optical pulse duration (cf. Methods). Thermal noise is intrinsically rejected in the cross-correlation process[25] but can be deduced from the measured microwave power to be -181 dBc.Hz$^{-1}$. Between 100 kHz and 1 MHz, we are limited by a servo bump due to the residual in-loop errors from the comb phase lock loop.

The measured absolute single-side-band (SSB) phase noise PSD for the 12 GHz microwave signal generated by our optical-frequency-comb frequency division scheme, obtained locking the two auxiliary microwave references and the DUT to three independent ultra-stable optical references, is displayed in Fig.3 (red curve). To our knowledge, this is the lowest absolute phase noise ever reported both close and far from the carrier[9]. At low offset Fourier frequencies, from 1 to 400 Hz, the phase noise is almost fully determined by the CW laser reference, indicating that we performed a close to complete transfer of frequency stability from optics to microwave. The -106 dBc.Hz$^{-1}$ at 1 Hz offset phase noise level is only 3 dB higher than what is inferred from the measured optical reference phase noise. The spurs in the Fourier frequency range from 10 Hz to 3 kHz originate from the 50 Hz power line harmonics due to imperfect electromagnetic shielding and acoustic noise coupling to the fiber-optics setup despite the laser being acoustically isolated and the fiber link from the CW laser to the comb system being noise-cancelled. Between 3 kHz and 1 MHz, the transfer is only limited by the residual phase noise of the optical frequency division scheme and the in-loop errors from the reference laser Pound-Drever-Hall lock.

Improvements to the close-to-carrier phase noise could be obtained with better ultra-stable laser frequency stability. Longer or cryogenic reference cavities[26, 27] with crystalline coatings[28] or spectral hole burning stabilization[29] could lower the phase noise limit by one order of magnitude. To be solely limited by the optical reference, shot noise could be reduced by using higher power-handling capabilities photodiode[30]. However, the most stringent requirement is photodiode flicker below the state-of-the-art -140 $f^{-1}$ dBc.Hz$^{-1}$ demonstrated here. Our extremely low noise setup is the ideal testbed for that development, offering the opportunity to study fundamental photodetection limits and suitable for any phase noise characterization, beyond comb-based systems.

Terabit communication systems with high-speed data transmission, high-stability fountain atomic clocks, very long baseline radio astronomy, as well as high-accuracy navigation and radar systems are direct applications that could benefit from the new level of microwave purity demonstrated in this paper. In particular, low phase noise in defense pulse-Doppler radar will enhance moving targets detection to an unprecedented level of resolution. Furthermore, this result paves the way to compact, robust, mobile microwave sources with ultra-low phase noise based on reliable technologies that have become readily accessible.

## Methods

**Optical frequency comb.** The fiber-based optical frequency comb (FOFC) consists of three parts: a femtosecond laser, an f-to-2f interferometer module, and an Erbium doped fiber amplifier (EDFA), all made with polarization-maintaining fiber. Based on the nonlinear amplifying loop mirror mode-locking principle, the laser reaches a steady state in a few seconds and can remain in the mode-locked state for several months. The repetition rate of the FOFC is coarsely tunable over 5 MHz around 250 MHz to accommodate the mode-spacing and then tune the frequency of the optically generated microwave signal. An intra-cavity electro-optical modulator (EOM) provides feedback on the comb repetition rate with a bandwidth around 1 MHz[31].

The laser output is pre-amplified up to 30 mW by a first EDFA and split in two parts used to seed the f-2f interferometer for CEO frequency detection and the high-power amplifier for microwave generation. The high-power EDFA boosts the optical power up to 350 mW and passively reduces the relative intensity noise (RIN) through seed saturation. It is pumped by three low-noise single-mode high-power laser diodes, all set to specific currents for avoiding parasitic mode-hopping. For each diode, the lowest RIN is observed when the current set-point is at equal distance from two successive mode-hops. The output shows a pulse width below 50 fs, with an optical spectrum broader than 60 nm FWHM.

**Ultra-stable laser and its phase noise characterization**: The reference CW laser is a 1542 nm semiconductor laser locked to an ultra-stable Fabry-Perot cavity (Q=380000) via Pound-Drever-Hall (PDH) techniques (500 kHz bandwidth), providing a fractional frequency stability of 5.5 x $10^{-16}$ at 1 s. The reference laser light is transferred to the FOFC using a 50-meter fiber link with an acousto-optic modulator (AOM)-based fiber noise canceller[32]. The laser phase noise is characterized by the same cross-correlation scheme that is used for the microwave system but operating on optical beatnotes rather than microwave mixed signals. Three 1542 nm ultra-stable lasers locked to three distinct cavities via PDH scheme were used. Each optical phase noise measurement sees two lasers act as references for characterizing the third one. The frequency difference between these three lasers is below 600 MHz.

**Optical frequency comb phase locking.** The optical phase locking technique of the FOFC to the ultra-stable reference laser is similar to what was reported previously[2,3,17]. Briefly, the booster amplifier output is followed by a fibered three-port thin-film optical add-drop multiplexer (OADM). The first output port delivers a signal filtered around 1542 nm with a bandwidth of 0.8 nm. At the second output, the unfiltered part of the initial spectrum is used for the microwave generation. The filtered signal is beat with the ultra-stable reference laser $\nu_{CW}$ leading to a beat-note signal $f_b = \nu_{cw} - Nf_r - f_0$. This beat-note signal is mixed with the CEO frequency, $f_0$ to get an $f_0$-independent signal at 880 MHz. This frequency is digitally divided by eight and compared with a fixed frequency from a direct digital synthesizer to generate an error signal. The error signal is then processed through a fast analog loop filter, and fed to the intra-cavity EOM and piezo-electric actuators so as to stabilize the repetition rate of the comb.

**High-linearity low-noise photodiode.** A top-illuminated dual-depletion region InGaAs/InP photodiode with optimized illumination profile was used for this work. The illumination profile is improved through beam-shaping *via* gradient index (GRIN) lens coupling. The power handling capability and linearity of the photodiode, which are limited by the collapse of the applied electric field due to photo-generated space charge effects, increase monotonically with its reverse bias. At 15 V reverse bias, the photodiode shows a linear response up to 4 Vpp output amplitude and deliv-ers a compressed output up to 6 Vpp. To avoid any thermally-induced device failure, the photodiode chip was integrated with a thermo-electric cooler and temperature sensor and assembled in a compact 8-pin microwave package to extract the heat from the chip and ensure reliable operation. In the final measurements presented here, the photodiode is typically illuminated with 15 mW of optical power, yielding 2.5 mW of 12 GHz RF power for a DC photocurrent of 8 mA. From the data shown in Fig.2 we can infer a flicker phase noise below -140 $f^{-1}$ dBc.Hz$^{-1}$, comparable to the best modified uni-travelling-carrier photodiodes[23].

**AM-PM conversion characterization and control.** In order to quantify and control the amplitude-to-phase noise conversion occurring through the photo-detection process, we implemented an FPGA-based system modulating the intensity of the FOFC output and coherently demodulating the relative phase between the photo-detected signal and a reference microwave oscillator. The intensity of the comb output light is modulated with an AOM placed right before the photodiode. The photodiode output signal is split between low frequencies (DC-30MHz) and high frequencies (30MHz-65GHz) *via* a high-quality diplexer. The DC part is digitally processed to extract the effective amplitude modulation. The AC part is filtered around 12 GHz with a narrow band cavity filter and directed to the RF port of a termination insensitive triple-balanced mixer ensuring broadband impedance matching at each port. The LO port of this mixer is fed with the 12.005 GHz signal from another stabilized FOFC acting as a low-noise microwave phase reference. The down-converted heterodyne signal (5 MHz) obtained at the IF port of the mixer is digitally phase-demodulated at the very frequency used for the amplitude modulation. The system allows slow active feedback for keeping the AM-PM conversion near a zero point.

The mixer input powers are +15dBm on the LO port and -5dBm on the RF port. The modulation frequency is conveniently chosen to be 1.23 kHz in order to avoid disturbance from parasitic power supply harmonics while preserving an efficient modulation.

**Dispersion Management.** When propagating through the 20 meters of optical fibers composing the setup, the femtosecond pulses spread and acquire considerable chirp resulting in an increased pulse duration. In order to reduce the pulse duration below the picosecond level, dispersion is compensated by inserting negative dispersion optical fibers (DCF38) at the output of the pulse rate multipliers. After theoretical estimation of the required fiber length, the optimization is done iteratively by steps of 2.5 cm while monitoring the pulse duration with commercial interferometric autocorrelator.

**Shot noise limit.** It has been previously reported that when photo-detecting a train of optical pulses, the shot noise limit is related to the optical pulse width[20]. Although we do observe and confirm such a dependency, the measured white phase noise floor of the microwave signal obtained illuminating the photodiode with 800 fs pulses deviates from the theoretical prediction by more than 20 dB (-196 dBc.Hz$^{-1}$). This significant discrepancy has been thought to originate from photocarrier scattering and absorption in the photodetector resulting in an increased timing jitter of the electrical pulse train[33].

**Phase noise characterization.** The 12 GHz harmonics from the FOFC under test is selected by a narrow filter (10 MHz bandwidth) and split in two parts. Each part is sent to the RF port of a termination insensitive triple-balanced mixer ensuring broadband impedance matching at each port and mixed with two distinct 12 GHz signals from two independently stabilized FOFC acting as references. The two reference signals (12.005 GHz and 11.995 GHz) are amplified to +13dBm by low phase noise narrow-band microwave amplifiers. 80 dB isolators are inserted at every mixer port in order to avoid any parasitic feedback that has been observed to undermine the phase noise result and the AM-PM conversion coefficient meas-urement. After the mixers, two heterodyne beat notes at 5 MHz are obtained and sent to the digital heterodyne cross-correlator[34]. Each heterodyne signal is sampled at 250 MS/s in an analog-to-digital converter clocked by a specific low-phase noise oven-controlled crystal oscillator slowly phase locked to a cryogenic sapphire oscillator for long-term stability (30 Hz bandwidth). The signal is digitally down-converted to DC and processed in an FPGA to yield in-phase and quadrature components down-sampled to 2 MS/s. The samples are transferred via Gbit-Ethernet to a computer where they are real-time analyzed by a Python-based software to obtain the phase noise information. The typical phase noise floor of the instrument for a 10 MHz carrier is -130 dBc.Hz$^{-1}$ at 1 Hz offset and below -180 dBc.Hz$^{-1}$ for Fourier frequencies beyond 1 kHz. Moreover the heterodyne nature of the cross-correlator renders it little sensitive to amplitude noise with more than 20 dB of amplitude-to-phase conversion rejection. The usual averaging time required to obtain the results displayed in Fig.2 is around 18 hours.

## Acknowledgements

We thank Jose Pinto for his help with the electronics and Rodolphe Le Targat for the reference laser distribution. This work is funded by the DARPA PULSE program (PureComb project), the FIRST-TF Labex and the Eurostar Eureka program (STAMIDOF project).

## Author contributions

X.X., R.B. and D.D. set up the experiment and carried out the measurements. M.G., W.H., M.L. and R.H. conceived and built the low-noise optical frequency combs. A.J. and S.D. provided the photodiodes. C.A. programmed the cross-correlator hardware. M.L. made the RF chains in the cross-correlator. P.A.T. and G.S. fabricated the pulse rate multipliers. X.X. and R.B. obtained the final data, prepared the manuscript and gathered the contributions from all the other co-authors. Y.L. designed the experiment and lead the project.

## Competing financial interests

The authors declare no competing financial interests.

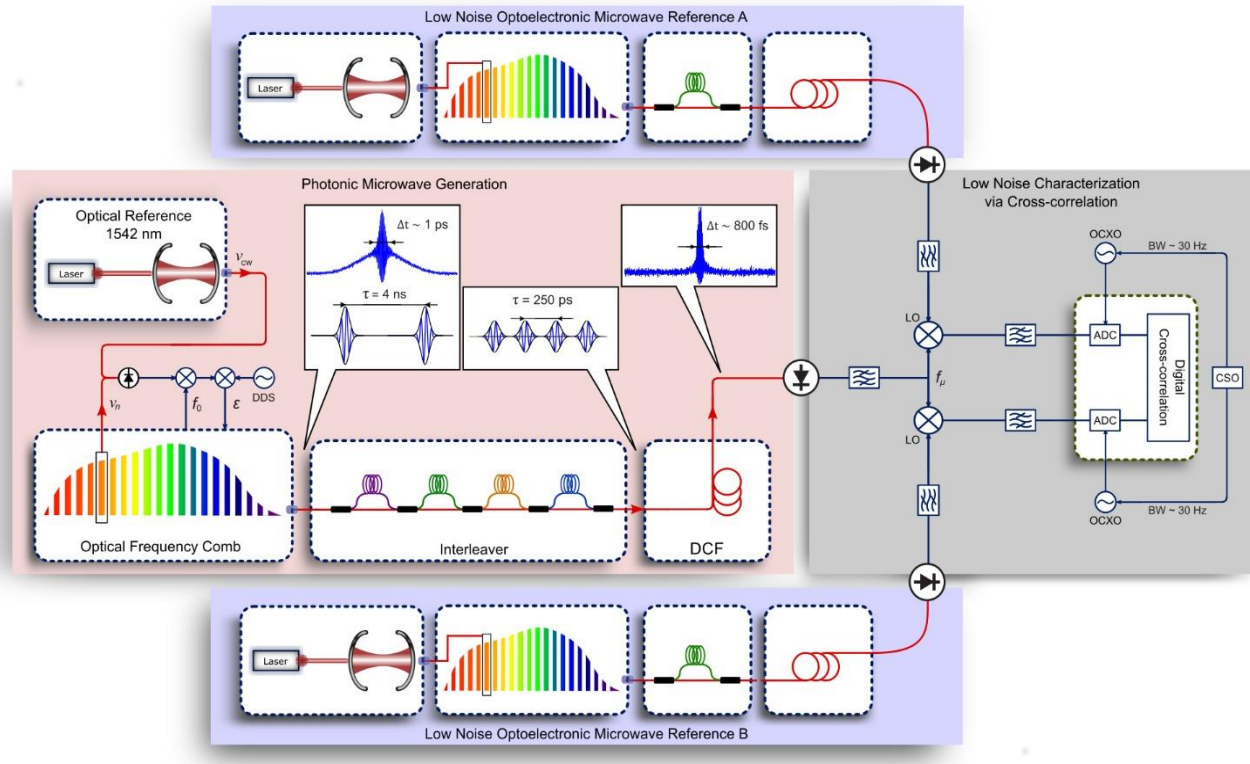

**Figure 1. Experimental setup for low-noise microwave generation and characterization.** DCF: dispersion compensation fiber, LO: local oscillator port, ADC: analog-to-digital converter, DDS: direct digital synthesizer, OCXO: oven-controlled crystal oscillator, CSO: cryogenic sapphire oscillator, BW: bandwidth. The left part of the setup is used for generating a record low noise 12 GHz microwave signal from an ultra-stable optical reference at 1542 nm *via* optical division. The right part is used for characterizing the microwave signal phase noise using a heterodyne digital cross-correlation method. The two auxiliary optoelectronic microwave references are also obtained by low noise optical division of the light from two additional distinct ultra-stable laser references at 1542 nm.

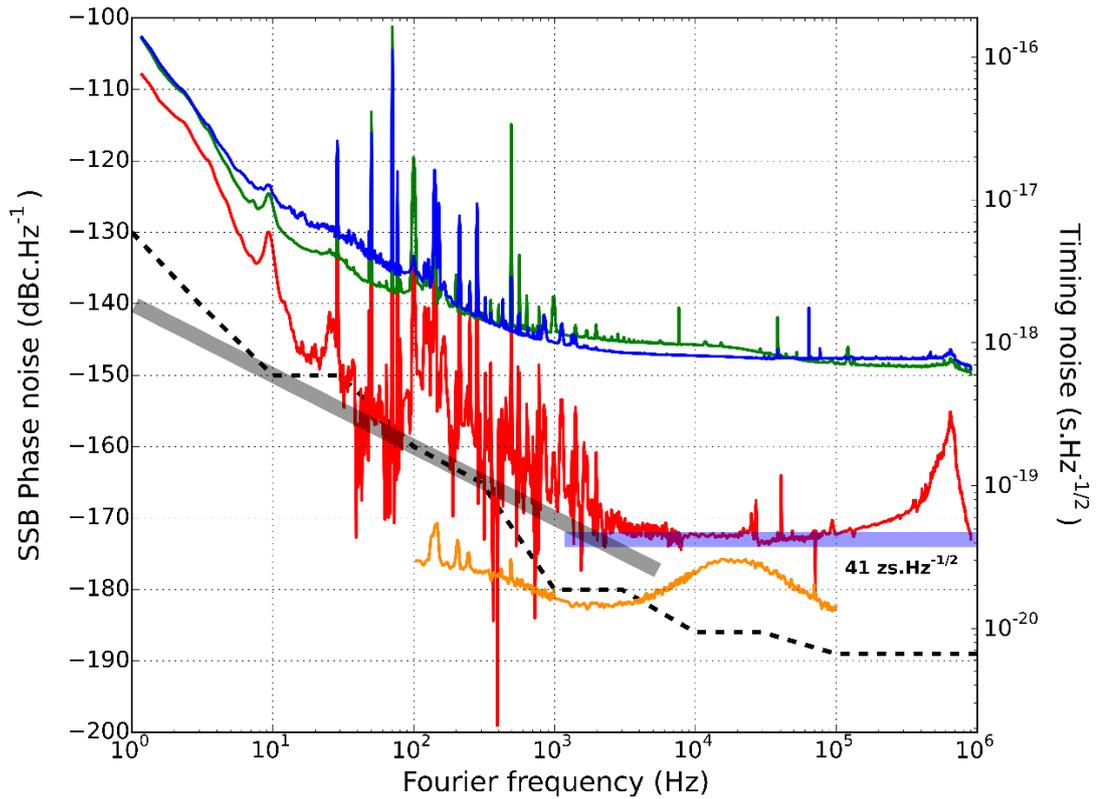

**Figure 2. Additive phase noise contribution of the frequency division scheme.** The green and blue lines are the phase noise of the 5 MHz RF signals obtained from the beat between the signal under test (at 12 GHz) and the auxiliary sources (at 11.995 GHz and 12.005 GHz). The phase data corresponding to these two curves are cross-correlated to yield the phase noise of the device under test (red line). In this measurement, all the lasers are stabilized using the same laser reference so that it represents the residual phase noise introduced in the optical division process and thus, the quality of the frequency stability transfer. It reaches a white timing noise floor of 41 zs.Hz$^{-1/2}$ (thick blue line). The outstanding photodiode flicker can be inferred to be below -140/f dBc.Hz$^{-1}$ (bold grey line). The orange line is the projected RIN-induced phase noise after 33dB of AM-PM conversion rejection. The dashed black line is the measurement noise floor.

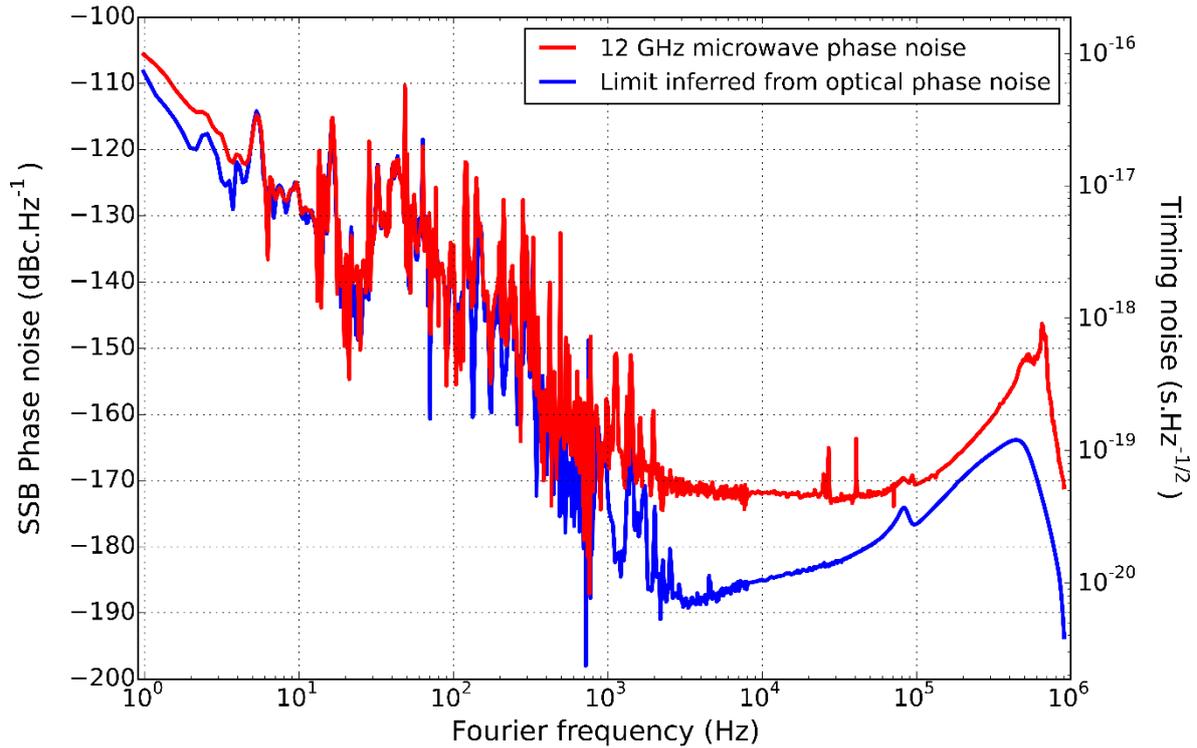

**Figure 3. Absolute transfer of spectral purity from optics to microwave.** The absolute phase noise of the 12 GHz microwave signal we generated (red line) is limited by the optical phase noise of the laser reference at 1542 nm (blue line) and by the residual in-loop error of the phase lock loop used to synchronize the repetition rate of the comb with the optical reference (bump above 700 kHz offset frequency). The phase noise of the 12 GHz signal sticks almost entirely to the optical phase noise, indicating a close to complete absolute transfer of purity from optical to microwave. Between 3 kHz and 100 kHz Fourier frequency, the transfer remains partial but offers an original opportunity to probe shot noise levels. The spurs ranging from 10 Hz to 1 kHz are due to acoustic noise picked up in the various fiber links and electromagnetic interference. The optical phase noise has been characterized independently by cross-correlation of RF optical beats from ultra-stable cavities.